\documentclass[preprint]{revtex4}%
\usepackage{amsmath,amssymb,amsthm}
\usepackage{graphics}
\usepackage{graphicx}
\usepackage{amsmath}
\usepackage{amsfonts}
\usepackage{amssymb}%
\providecommand{\U}[1]{\protect\rule{.1in}{.1in}}
\begin{document}
\title{Effective Field Theory for Light in Disordered Atomic Medium}
\author{M.B. Smirnov}
\affiliation{\;RRC Kurchatov Institute, Kurchatov sq. 1. 123182
Moscow, Russian Federation}
\author{M.A. Baranov}
\affiliation{\;RRC Kurchatov Institute, Kurchatov sq. 1. 123182
Moscow, Russian Federation} \affiliation{Van der Waals-Zeeman
Instituut, Unversiteit van Amsterdam, Valckenierstraat 65, 1018 XE
Amsterdam, The Netherland}

\begin{abstract}
We develop a field theory approach to light propagation in a gas of resonant
atoms taking into account vector character of light and atom-atom
interactions. Within this approach, we calculate the propagator of the
electric field for both short and long-range density-density correlation
functions of the gas.

\end{abstract}
\maketitle

\section{Introduction}

Transport of waves through strongly scattering disordered media has received
much attention during past years. The character of this process is determined
by the multiply elastic scattering of light. In a sample of randomly
distributed scatterers, the initial direction of the wave is fully randomized
by multiple scattering, and a diffusion picture seems to be an appropriate
description of light propagation in the case of small densities of scatters
\cite{Hulst80Multiplyscattering}. Accordance to this picture, he transmission
coefficient of a medium is inversely proportional to the sample thickness,
i.e. follows the familiar Ohm's law. Despite successful predictions, this
theory does not take into account interference effects. P.W. Anderson was the
first\ to demonstrate that sufficiently strong disorder can result in a
localization of quantum-mechanical wave functions \cite{PR58Anderson}. This
phenomenon is referred to as Anderson localization. In fact, even away from
the localized phase, in the so-called weak localization regime, wave
interference affects the physics of light propagation. For example, the
constructive interference of counter-propagating light amplitudes results in
the coherent backscattering peak \cite{JPhFr88Akkermans,PRB88Lagenijk}. In the
past 20 years, mesoscopic physics has been developed, with beautiful
experimental and theoretical results pertaining to the weak localization
regime \cite{PR96Lagendijk,PR99Nieuwenhuizen,PR98Lagendijk}.

A system of randomly distributed atoms with an atomic transition frequency
close to the light frequency, provides necessary conditions for experimental
observation of backscattering
\cite{Condmat06Miniatura,PRL99Miniatura,PRA01Miniatura,PhysLetA94Gora}. In
fact, under these conditions, light scattering on an individual atom is
characterized by a large cross-section. However, in contrast to classical
scatters, in this system one has to take into account resonance dipole-dipole
interactions that alter the character of the light propagation. There are two
different mechanisms of excitation transfer. According to the first mechanism,
an excited atom decays into the ground state by emitting a photon. Then the
photon propagating through a system of resonance scatters, is absorbed by
another atom, and thus excitation transfer takes place. In the second
mechanism, an excitation may transfer from atom to atom nonradiatively by the
resonance dipole-dipole interaction. This is equivalent to including the
longitudinal component of the electromagnetic field. This mechanism becomes
more important with increasing the density of atoms $n$. If the average number
of atoms per cube with the size of the wavelength $\lambda$ exceeds unity,
$n\lambda^{3}\gtrsim1$, the typical energy of dipole-dipole interactions is
comparable with the inverse of the the radiative lifetime of excited atoms,
and the transport of nonradiative photons dominates the transport of radiative
ones. The resonance dipole-dipole interaction also plays an important role in
the radiative transfer by binary collisions of atoms in
gases\cite{PRL98MourachkoRydbergGas,PRA04MourachkoRydGas}.

This problem has been actively studied under in the case of low atom number
densities\cite{JPA02Miniatura,PRA01Miniatura,PRL05Muller}. By using the scalar
model of light it has been demonstrated that interference effects are
substantially suppressed as a result of the motion of
scatters\cite{JEPT84Golub}. In Refs. \cite{PhysLetA94Gora} the authors
considered the vector character of light and took into account atom-atom
interaction. It was found that close to the resonance, the transport slows
down considerably because the light is captured by atoms for a longer time.
Such a behavior results in a considerable reduction of the speed of light.
This effect have been observed for the classical systems, but for the dipolar
gas the effect differs\cite{Condmat06Miniatura,PRL99Miniatura}.

In this paper we focus our attention on the case $n\lambda^{3}\ll1$ and
consider different density-density correlation functions for light scatters.
The latter will be treated as a ($1+3$)-level atom (In other words, we
consider atoms with a resonant $s-p$ transition.). We also take into account
the effects of dipolar atom-atom interactions on the light propagation. The
influence of the atomic density-density correlation function ( the second
cumulant, or second central moment of the probability distribution for
different atomic realizations) is of special interest for us. It appears that
in the considered case of a low density system, only this correlation function
affects the distribution of total transmission coefficient. We therefore
assume that the other cumulants are zero (that is we consider gaussian
probability distributions). This means that a general high order density
correlation function reduces to the product of pair density-density correlations.

The paper is organized as follows. In Sec. II, we introduce the action for the
($1+3$)-level dipolar gas interacting with quantum electromagnetic field. An
effective action for the light propagating in the gas, is derived in Sec. III.
We evaluate the propagator for the electric field and identify the
localization length of light in the gas in Sec. IV. There we also present
simple relationship between the density-density correlation function and the
light propagator, and analyze the result for various correlations in the gas.
Finally, in Sec. V we conclude.

Throughout the paper we adopt the system of units in which the speed of light
$c$, the electron charge $e$, and the Planck constant $\hbar$ are set to
unity, $\ c=e=\hbar=1$.

\section{Field theory description of the system}

\subsection{General considerations}

The propagation of light in a medium is described by the Maxwell equations
supplemented with ones characterizing the properties of the medium. We express
the solution in terms of propagators that relate the amplitude of the
electromagnetic 4-vector potential $A_{k}(X)$ ($k=0,1,2,3$) at a space-time
position $X={t,\mathbf{x}}$, with $A_{n}(X^{\prime})$ and is defined as
\begin{equation}
D_{kn}(X-X^{\prime})=\langle G|TA_{k}(X)A_{n}(X^{\prime})|G\rangle
\end{equation}
where $\left\vert G\right\rangle $ is the ground state of the system, when the
all atoms in the ground states and the photons are absent. We have used above
the fact that the medium is transactionally invariant. For the medium that
consists of scatters at random positions, the propagator of a given sample
depends on the location of scatterers. To get general characteristics for
light propagation in the disorder medium one have to average the propagator
over disorder:
\begin{equation}
\bar{D}_{kn}(X-X^{\prime})=\langle\langle G|TA_{k}(X)A_{n}(X^{\prime
})|G\rangle\rangle_{dis} \label{eqGreen}%
\end{equation}
The averaged propagator is referred to as the amplitude propagators.

For further calculation the quantum field theory methods are applied. Thus,
for the evaluation the expression (\ref{eqGreen}) we have to construct the
action for the light interacting with dipole gas. The action for the system
depends on atoms degrees of freedom as well as electromagnetic field one. In a
general form it can be written as
\begin{equation}
S=S(A_{k},\psi_{i}^{n},\bar{\psi}_{i}^{n})
\end{equation}
where $\psi_{i}^{n}$ are fields corresponds to atoms and $A_{k}$ 4-vector of
electromagnetic field. Therefore the partition function $Z$ for the system is
\begin{equation}
Z=\int\prod\limits_{k=0,1,2,3}\mathcal{D}A_{k}\prod\mathcal{D}\left\{
\psi_{i}^{n},\bar{\psi}_{i}^{n}\right\}  \exp\left[  iS(A_{k},\psi_{i}%
^{n},\bar{\psi}_{i}^{n})\right]  \label{eqZ1}%
\end{equation}
On the base of the participation function (\ref{eqZ1}) we can evaluate the
properties of a certain realization of disorder. To find the averaged
propagators (\ref{eqGreen}) one have averaged over disorder. Applying
continual approach we figure distribution for realization of the number
density $\rho(x)$ as:
\begin{equation}
P[\rho(x,t)]=C\exp\left[  -\sum\limits^{n}\int dx_{1}\ldots dx_{n}\frac
{\rho(x_{1}t_{1})\rho(x_{2}t_{2})\cdot\ldots\cdot\rho(x_{n}t_{n})}%
{n!\Gamma^{(n)}(x_{1}\ldots x_{n},t_{1}\ldots t_{n})}\right]
\label{eqDistrib}%
\end{equation}
Here $\Gamma^{(n)}(x_{1}\ldots x_{n},t_{1}\ldots t_{n})$ is n-cumulant for the
distribution. The first cumulant is equal to the averaged number density
$\rho_{0}(x,t)$, while the second one coincides with dispersion for the number
density :
\begin{equation}
\Gamma^{(2)}(x_{1}t_{1},x_{2}t_{2})=\left\langle (\rho(x_{1}t_{1})-\rho
_{0}(x_{1}t_{1}))(\rho(x_{2}t_{2})-\rho_{0}(x_{2}t_{2}))\right\rangle
\end{equation}
If the number density of scatters is not high, the other cumulant are close to
zero. For further evaluation we take into account that the number density can
evolve with time, then the improved distribution (\ref{eqDistrib}) for
$\rho(x,t)$ can be depicted as:
\begin{equation}
P[\rho(x,t)]\sim\exp\left[  -\frac{1}{2}\int dxdx^{\prime}\frac{\left(
\rho(x,t)-\rho_{0}(x,t)\right)  \left(  \rho(x^{\prime},t^{\prime})-\rho
_{0}(x^{\prime},t^{\prime})\right)  }{\Gamma^{(2)}(x,t;x^{\prime},t^{\prime}%
)}\right]
\end{equation}
There the averaging-out over disorder realization results in appearing a
supplementary field $\rho(x,t)$
\begin{equation}
Z=\int\prod\limits_{k=0,1,2,3}\mathcal{D}A_{k}\prod\mathcal{D}\left\{
\psi_{i}^{n},\bar{\psi}_{i}^{n}\right\}  \mathcal{D}\rho(x,t)\;P[\rho
(x,t)]\exp\left[  iS(A_{k},\psi_{i}^{n},\bar{\psi}_{i}^{n})\right]
\end{equation}

\subsection{Model for atoms.}

For the evaluation the field propagator we are in need of the full action for
the system $S(A_{k},\psi_{i}^{n},\bar{\psi}_{i}^{n})$. We start the analysis
from consideration of a single atom immersed into an external electromagnetic
field. The total action for a single atom $S_{tot\_atom}$ in the
electromagnetic field consists of three terms: the action of the free atom
$S_{atom}$, the free electromagnetic field $S_{f}$ and the term caused by the
interaction of the electromagnetic field with the atom $S_{int}$.

Let us find the action of a free atom. All evaluation will be carried out in
the assumptions, that the atom is infinitively heavy. It allows us to take
into account the evolution of just internal degrees of freedom. The
translation degrees of freedom of the atom are not taken into account, because
the effect of the momentum exchange between the atom and incident
electromagnetic field is supposed to be small. In general case atomic
structure is fairly complex and with increasing the number of electrons in the
atom its complexity is increasing and the number of energy states grows
\cite{Landau3,Sobelman1979}. The action of the external electromagnetic field
inspires transitions between the atomic states. If the atom with ground state
energy $\varepsilon_{g}$ and excited state energy $\varepsilon_{a}$ is
subjected to the monochromatic radiation with frequency $\omega$ and the
conditions $\omega\approx|\varepsilon_{a}-\varepsilon_{g}|$ holds true, the
most of transitions occurs between the ground and the excited state. The
others states does not affect on the atom dynamics in the electromagnetic
field. It is assumed the character of the resonance transitions to dipole. In
accordance with the selection rules for the dipole transitions, the final
orbital momentum for an atom $L_{f}$ and initial orbital momentum $L_{i}$ are
related by $|L_{i}-L_{f}|=1$. Thus, if the ground state is singlet, the
excited state has to be triplet, then transitions takes between four states
$|E,l\rangle$ with the energy $E$ and projection of orbital momentum $l$
\begin{equation}
\begin{array}{*{20}c} {|g\rangle=|\varepsilon_{g},l=0\rangle} \\ {|-\rangle=|\varepsilon_{a},l=-1\rangle} \\ {|0\rangle=|\varepsilon_{a},l=0\rangle} \\ {|+\rangle=|\varepsilon_{a},l=+1\rangle} \\ \end{array}
\end{equation}
So consider the system that contains atoms in the lowest ($E=\varepsilon
_{g},l=0$) and first excited states ($E=\varepsilon_{a},l=-1,0,1$) where we
let $c_{i}^{+}$ with $i=g,-,0,+$: create these respective states.If
$|Vac\rangle$ is an atom vacuum state (i.e. the state that corresponds to the
atom absence), these operators satisfy the following condition:
\begin{equation}
\begin{array}{*{20}c} {|i\rangle=c_{i}^{+}|Vac\rangle \qquad |Vac\rangle=c_{i}|i\rangle} \qquad i=-1,0,1,g \\ {\left\{c^{+}_{i},c_{k}\right\}=\delta_{ik}\qquad \left\{c^{+}_{i},c^{+}_{k}\right\}=0, \qquad \left\{c_{i},c_{k}\right\}=0, \;\forall i,k } \\ \end{array}
\end{equation}
We will use further the change of basis $c_{g}^{+},c_{x}^{+}=\left(
{c_{-}^{+}+c_{+}^{+}}\right)  /\sqrt{2}{\ },c_{y}=-i\left(  {c_{+}^{+}%
-c_{-}^{+}}\right)  /\sqrt{2}{\ },c_{z}^{+}=c_{0}^{+}$, and the hamiltonian of
a free atom can be described in new terms of annihilation and creation
operators $c_{i},c_{i}^{+}$ where $i=g,x,y,z$ as%

\begin{equation}
\hat{H}=\varepsilon_{a}\hat{c}_{x}^{+}\hat{c}_{x}+\varepsilon_{a}\hat{c}%
_{y}^{+}\hat{c}_{y}+\varepsilon_{a}\hat{c}_{z}^{+}\hat{c}_{z}+\varepsilon
_{g}\hat{c}_{g}^{+}\hat{c}_{g} \label{eq1}%
\end{equation}
Let us introduce the Grassmann fields $\psi_{i}$ and $\bar{\psi}_{i}$ that
correspond to annihilation and creation operators $c_{i},c_{i}^{+}$,
$i=g,x,y,z$. If the fields $\psi_{i}$ and $\bar{\psi}_{i}$ correspond to
coordinate and momentum, then the free atom Lagrangian takes the form:
\begin{equation}
L_{atom}=\sum\limits_{i=g,x,y,z}{\bar{\psi}_{i}(t)i\partial_{t}\psi_{i}%
(t)}-\varepsilon_{g}\bar{\psi}_{g}(t)\psi_{g}(t)-\varepsilon_{a}%
\sum\limits_{\alpha=x,y,z}{\bar{\psi}_{a}(t)\psi_{a}(t)}
\label{Lag_singleatom}%
\end{equation}
and the action for a single atom
\begin{equation}
S_{atom}=\int dt\left[  \sum\limits_{i=g,x,y,z}{\bar{\psi}_{i}(t)i\partial
_{t}\psi_{i}(t)}-\varepsilon_{g}\bar{\psi}_{g}(t)\psi_{g}(t)-\varepsilon
_{a}\sum\limits_{\alpha=x,y,z}{\bar{\psi}_{a}(t)\psi_{a}(t)+}i\delta
\sum\limits_{i=g,x,y,z}{\bar{\psi}_{i}(t)\psi_{i}(t)}\right]
\label{Ac_singleatom}%
\end{equation}
A short imaginary parts appears to satisfy causality principle for atom
evolution.Taking the energy of the ground state is equal to zero we can
rewrite action for a single atom as
\begin{equation}
S_{atom}=\int\bar{\psi}_{g}(t)\left[  i\partial_{t}+i\delta\right]
\psi _{g}(t)dt+\int\bar{\psi}_{\alpha}(t)\left[
i\partial_{t}-\Delta
+i\delta\right]  \psi_{\alpha}(t)dt,\Delta=\varepsilon_{a}-\varepsilon_{g}%
\label{Ac_singleatom_fin}%
\end{equation}

\subsection{Interaction of atoms with light}

The other two terms of the total action $\left(  S_{tot\_atom}\right)  $
$S_{f}$ and $S_{int}$ depend on the properties of electromagnetic field. If
the electromagnetic field is determined by 4-potential $A_{i}$, the action
associated with the free e-m field\cite{Landau4,Peskin1995} is
\begin{equation}
S_{f}=-\frac{1}{4\pi c}\int d^{4}x\left(  \frac{1}{4}F_{ik}F^{ik}+\frac
{1}{2\xi}(\partial^{k}A_{k})^{2}\right)  \qquad F_{ik}=\frac{\partial A_{k}%
}{\partial x^{i}}-\frac{\partial A_{i}}{\partial x^{k}} \label{Action_field}%
\end{equation}
The constant $\xi$ appears as result of gauge invariance for the
electromagnetic fields\cite{Landau4,Peskin1995}. Further we choose an
appropriate gauge for evaluations. The term resulted from the interaction has
to consists of field variable $A_{i}$ as well as 4-current $j^{i}$, which is
determined by atom properties:
\begin{equation}
S_{int}=-\frac{1}{c^{2}}\int{d^{4}xA_{i}j^{i}}
\label{eq_action_for_interaction}%
\end{equation}
If the atom is placed at the point $\mathbf{x_{n}}$ and the wavelength of
electromagnetic field surpasses remarkably its size, we can treat the atom as
point dipole and the 4-current produced by it is
\begin{equation}
j^{n}(xt)=\left\{
{\begin{array}{*{20}c} {c\rho (xt)} \\ {\vec j(xt)} \\ \end{array}}\right\}
=\left\{
{\begin{array}{*{20}c} { - cd_i (t)\partial _i \delta (\vec x - \vec x_a )} \\ {\dot d_i (t)\delta (\vec x - \vec x_a )} \\ \end{array}}%
\right\}  \label{4current}%
\end{equation}
Integrating expression (\ref{eq_action_for_interaction}) over time and
applying for the 4-current (\ref{4current}), we get :
\begin{equation}
S_{int}^{n}=-\frac{1}{c^{2}}\int{d^{4}xA_{i}j^{i}}=\frac{1}{c}\int{E_{\alpha
}(x_{n},t)d_{\alpha}(t)dt}\qquad\alpha=1,2,3 \label{eq_action_field}%
\end{equation}
where $E_{\alpha}(x,t)$ is electric field strength:
\begin{equation}
E_{\alpha}(x,t)=-\left[  {\partial_{\alpha}A_{0}(x,t)+\partial_{0}A_{\alpha
}(x,t)}\right]
\end{equation}
If $\ d_{0}$ is the value of dipole moment associated with transition
$\varepsilon_{g}\rightarrow\varepsilon_{a}$, the component of the operator
${d_{\alpha}(t)}$ is written as%

\[
{d_{\alpha}(t)=d_{0}\left[  {\bar{\psi}_{\alpha}(t){\psi_{g}(t)}+\bar{\psi
}_{g}(t)\psi_{\alpha}(t)}\right]  }%
\]
The expression (\ref{eq_action_field}) is exact for point-like dipole.

Thus the total action for the system of the atom and the electromagnetic field:%

\begin{equation}%
\begin{array}
[c]{l}%
S_{tot_atom}=-\frac{1}{4\pi c}\int d^{4}x\left(  \frac{1}{4}F_{ik}%
F^{ik}+\frac{1}{2\xi}(\partial^{k}A_{k})^{2}\right) \\
\frac{1}{c}\int{dtE}_{\alpha}({x_{a},t}){d_{0}\left[  {\bar{\psi}_{\alpha
}(t){\psi_{g}(t)}+\bar{\psi}_{g}(t)\psi_{\alpha}(t)}\right]  }\\
\int dt\bar{\psi}_{g}(t)\left[  i\partial_{t}+i\delta\right]  \psi
_{g}(t)dt+\int\bar{\psi}_{\alpha}(t)\left[  i\partial_{t}-\Delta
+i\delta\right]  \psi_{\alpha}(t)
\end{array}
\label{eq_atom_full_action}
\end{equation}
The total action of the atoms ensemble $S_{total}$ and the
electromagnetic field can be found by the generalizing
(\ref{eq_atom_full_action}). Naively one may assume the total
action for an atomic ensemble is sums actions of free atoms and
interaction terms over all ensemble, and the free electromagnetic
field term. Applying expression for $S_{atom}^{n}$ the action of a
single atom at a point $x_{n}$ (\ref{Ac_singleatom_fin})  and its
current $j^{i}(\mathbf{x_{n}})$ (\ref{4current}):
\begin{equation}
S_{total}^{\prime}=\sum\limits_{n}S_{atom}^{n}-\frac{1}{c}\int A_{i}%
\sum\limits_{n}j^{i}(\mathbf{x_{n}})+S_{f}\left(  A_{i}\right)
\end{equation}
But this naive approach leaves out of account the interactions between atoms,
resulted in an additional term in the total action. The force acted between
two atoms is referred to as the Van der Waals force. The elementary processes
on which the Van der Waals interaction is based is the exchange of the pair
photons between the two atoms. This force depends on the distance as well as
energy of resonance transitions. If the distance between the dipoles $R$ is
short, the virtual photons propagate instantly and the force is proportional
to $R^{-7}$. In the opposite case, the delay in propagation of photons should
be taken into account. The character dependence force on the distance is
changed and has a form $R^{-8}$. The interaction between dipoles has
electromagnetic nature. Thus the electromagnetic filed in the dipole gas
consists of two parts: the first one originates from the external source
$A_{i}^{ext}$ and the second part appears as a result of interaction between
dipoles $A_{i}^{qn}$. Further we are refereed to $A_{i}^{ext},A_{i}^{qn}$ as
"external" and "quantum" fields. Therefore the total action for dipole gas in
the electromagnetic field takes the form:%
\begin{equation}
S_{total}=\sum\limits_{n}S_{atom}^{n}-\frac{1}{c}\int\left(  A_{i}^{ext}%
+A_{i}^{qn}\right)  \sum\limits_{n}j^{i}(\mathbf{x_{n}})+S_{f}\left(
A_{i}^{qn}\right)  +S_{f}\left(  A_{i}^{ext}\right)  \label{TotalAction}%
\end{equation}
If the external field is equal to zero, one can integrates over the quantum
field $\ A_{q}$ and the new effective action is determined by the atomic
freedom of degree. This action includes sum free atomic action for all atoms
\ref{Ac_singleatom} and the term $S_{AA}$ related with interquartile
interaction
\begin{equation}
S_{AA}=-\frac{1}{{c^{3}}}\int{d^{4}xd^{4}x^{\prime}\frac{{J_{m}(x)J^{m}%
(x^{\prime})}}{{\left\vert {\overline{x}-\overline{x^{\prime}}}\right\vert }%
}\delta(\tau-\tau^{\prime}-|\overline{x}-\overline{x^{\prime}}|)} \label{Saa}%
\end{equation}
where $J^{m}=\sum\limits_{n}j^{m}(\mathbf{x_{n}})$ is the total 4-current
caused by all dipoles. If we consider the limit case, when the interaction can
be treated as instant, i.e. $\Delta\cdot R_{mn}/c\ll1$ ($R_{mn}$ is distance
between atoms n and m) the interaction takes the form:%

\begin{align*}
S_{AA}  &  =-\frac{1}{2}%
{\displaystyle\sum\limits_{m,n,m\neq n}}
{\displaystyle\int}
H_{mn}(t)dt\\
H_{mn}\left(  t\right)   &  =\frac{d_{\alpha}^{n}\left(  t\right)  d_{\alpha
}^{n}\left(  t\right)  R_{mn}^{2}-3d_{\alpha}^{n}\left(  t\right)  d_{\beta
}^{m}\left(  t\right)  R_{\alpha,mn}R_{\beta,mn}}{R_{mn}^{5}}%
\end{align*}
i.e. we get the dipole-dipole interaction for static limit, as we should. The
total action $S_{total}$ for the dipole gas is given by sum\ref{TotalAction}
with $j^{i}(\mathbf{x_{n}})$ determined by \ref{4current}.

Because the ground state of the system can be determined by as product of the
atomic the ensemble ground state and the electric field ground state
$\left\vert 0\right\rangle $(the state when the number of photons is equal to
0), the ground of the system is
\[
\left\vert G\right\rangle =\left\vert 0\right\rangle
{\displaystyle\prod\limits_{n}}
\left\vert g\right\rangle _{n}=\left\vert 0\right\rangle
{\displaystyle\prod\limits_{n}}
\bar{\psi}_{g}\left\vert Vac\right\rangle _{n}%
\]
the partition function for the ensemble is%

\begin{align*}
Z  &  =\lim_{t_{i}\rightarrow-\infty}\lim_{t_{f}\rightarrow\infty}%
\mathcal{N}\int\mathcal{D}\left\{  {A^{ext,0}A^{ext,1}A^{ext,2}A^{ext,3}%
}\right\}  \mathcal{D}\left\{  {A^{qn,0}A^{qn,1}A^{qn,2}A^{qn,3}}\right\} \\
&  \cdot\prod\limits_{n}\prod\limits_{i=g,1,2,3}\mathcal{D}\psi_{i}%
^{n}\mathcal{D}\bar{\psi}_{i}^{n}\psi_{g}^{n}\left(  t_{f}\right)  \bar{\psi
}_{g}^{n}\left(  t_{i}\right)  \cdot\exp[iS_{total}]
\end{align*}
here $\mathcal{N}$ is fields independent normalization constant.

\section{Effective field theory for light}

Above we have found the partition function for the interacting atomic
ensemble. The paper object is to derive the Greens function for
electromagnetic field. To carry out this deducing, we have to integrate over
atomic fields $\psi_{i}^{n}$. The integration over atomic variables in the
partitions function for the interacting dipole gas can not be performed
accurately and some approximations must be employed. For weak interaction,
perturbation theory can be applied, when the interacting term $\exp\left(
iS_{ints}\right)  $ is expanded in the form of a power series:%

\[
\exp\left(  iS_{ints}\right)  =1-\frac{S_{int}^{n2}}{2}+\frac{S_{int}^{n4}%
}{24}+\ldots
\]
we keep the terms up to the forth order. For brevity it is convenient to
introduce the following notation for the fields:
\begin{align*}
A^{i}  &  =\frac{\left(  {A^{qn}}\right)  ^{i}{+}\left(  {A}^{ext}\right)
^{i}}{\sqrt{2}}\\
B^{i}  &  =\frac{\left(  {A^{qn}}\right)  ^{i}{-}\left(  {A}^{ext}\right)
^{i}}{\sqrt{2}}%
\end{align*}
Then the partition function for the dipole gas is%

\begin{align*}
Z  &  =Z_{1}\cdot Z_{2}\\
Z_{n}  &  =\lim_{t_{i}\rightarrow-\infty}\lim_{t_{f}\rightarrow\infty
}\mathcal{N}_{n}\int\prod\limits_{i=g,1,2,3}\left[  \mathcal{D}\psi_{i}%
^{n}\mathcal{D}\bar{\psi}_{i}^{n}\right]  \psi_{g}^{n}\left(  t_{f}\right)
\bar{\psi}_{g}^{n}\left(  t_{i}\right)  \left(  1-\frac{S_{int}^{n2}}{2}%
+\frac{S_{int}^{n4}}{24}\right)  \cdot\exp[iS_{atom}^{n}]\\
Z_{1}  &  =\mathcal{N}_{A}\int\mathcal{D}\left\{  {A^{0}A^{1}A^{2}A^{3}%
}\right\}  \exp[iS_{field}^{1}]\prod\limits_{n}Z_{n}\\
Z_{2}  &  =\mathcal{N}_{B}\int\mathcal{D}\left\{  B{^{0}B^{1}B^{2}B^{3}%
}\right\}  \exp[iS_{field}^{2}]
\end{align*}
where%

\begin{equation}%
\begin{array}
[c]{l}%
S_{field}^{1}=\frac{1}{4\pi c}\int d^{4}x\left[  \frac{1}{4}F_{im}F^{im}%
+\frac{1}{2\xi}(\partial^{k}A_{k})^{2}\right]  \qquad F_{ik}=\frac{\partial
A_{k}}{\partial x^{i}}-\frac{\partial A_{i}}{\partial x^{k}}\\
S_{field}^{2}=\frac{1}{4\pi c}\int d^{4}x\left[  \frac{1}{4}P_{im}P^{im}%
+\frac{1}{2\xi}(\partial^{k}B_{k})^{2}\right]  \qquad P_{ik}=\frac{\partial
B_{k}}{\partial x^{i}}-\frac{\partial B_{i}}{\partial x^{k}}\\
S_{int}^{n}=\frac{\sqrt{2}{d_{0}}}{c}\int{d\tau\cdot E_{\alpha}(x\tau){\left[
{\bar{\psi}_{\alpha}^{n}(\tau)\psi_{g}^{n}(\tau)+\bar{\psi}_{g}^{n}(\tau
)\psi_{\alpha}^{n}(\tau)}\right]  }}\\
S_{atom}^{n}=\int{d\tau}\left[  {\sum\limits_{\alpha=1}^{3}{\bar{\psi}%
_{\alpha}^{n}(\tau)\left(  {i\partial_{\tau}-\Delta+i\delta}\right)
\psi_{\alpha}^{n}(\tau)}+\bar{\psi}_{g}^{n}(\tau)\left(  {i\partial_{\tau
}-i\delta}\right)  \psi_{g}^{n}(\tau)}\right] \\
E_{\alpha}(x_{n},\tau)=-\left(  \partial_{\alpha}A_{0}(x_{n},\tau)+{\partial
}_{0}{A_{\alpha}(x_{n},\tau)}\right)
\end{array}
\end{equation}
and $\mathcal{N}_{A},\mathcal{N}_{B},\mathcal{N}_{n}$ are constants. Because
the atomic fields do not mix in full action, the integration for each atom can
be performed independently. As result of integration we get finally for
$Z_{n}$
\begin{equation}
Z_{n}=\mathcal{\breve{N}}_{n}\left(
\begin{array}
[c]{c}%
1-i\frac{2{d}_{0}^{2}}{c^{2}}\int dt_{1}dt_{2}{{{{E_{a}(x_{n},t}}_{1}{)}%
E_{a}(x_{n},t}}_{2}{)\cdot G}_{e}\left(  t_{1}-t_{2}\right) \\
-i\left(  \frac{2{d}_{0}^{2}}{c^{2}}\right)  ^{2}\int dt_{1}dt_{2}dt_{3}%
dt_{4}\cdot{G}_{e}\left(  t_{1}-t_{2}\right)  {G}_{g}\left(  t_{2}%
-t_{3}\right)  {G}_{e}\left(  t_{3}-t_{4}\right) \\
\cdot{{{{E_{a}(x_{n},t}}_{1}{)}E_{a}(x_{n},t}}_{2}{{){{E_{b}(x_{n},t}}%
_{3}{{){{E_{b}(x_{n},t}}_{4}{)}}}}}%
\end{array}
\right)
\end{equation}
where the atomic Greens function for the ground $G_{g}\left(  t_{2}%
-t_{1}\right)  $ and the excited states $G_{e}\left(  t_{2}-t_{1}\right)  $
are defined as:
\begin{align*}
\langle\psi_{g}^{n}(t_{2})\bar{\psi}_{g}^{n}(t_{1})\rangle &  =iG_{g}\left(
t_{2}-t_{1}\right)  =\theta\left(  t_{2}-t_{1}\right) \\
\langle\psi_{a}^{n}(t_{2})\bar{\psi}_{b}^{n}(t_{1})\rangle &  =i\delta
_{ab}G_{e}\left(  t_{2}-t_{1}\right)  =\theta\left(  t_{2}-t_{1}\right)
\exp\left[  -i\Delta\left(  t_{2}-t_{1}\right)  \right]  \delta_{ab}%
\end{align*}
The final expression for $\prod\limits_{n}Z_{n}$ up to the forth order for
$d_{0}$ is%

\begin{equation}
\prod\limits_{n}Z_{n}=\mathcal{\breve{N}}\left(
\begin{array}
[c]{c}%
1-%
{\displaystyle\sum\limits_{n}}
i\frac{2{d}_{0}^{2}}{c^{2}}\int dt_{1}dt_{2}{{{{E_{a}(x_{n},t}}_{1}{)}%
E_{a}(x_{n},t}}_{2}{)\cdot G}_{e}\left(  t_{1}-t_{2}\right) \\
+\frac{1}{2}%
{\displaystyle\sum\limits_{n,m,n\neq m}}
\left(  -i\frac{2{d}_{0}^{2}}{c^{2}}\right)  ^{2}\int dt_{1}dt_{2}dt_{3}%
dt_{4}{\cdot G}_{e}\left(  t_{1}-t_{2}\right)  {G}_{e}\left(  t_{3}%
-t_{4}\right) \\
{{{{E_{a}(x_{n},t}}_{1}{)}E_{a}(x_{n},t}}_{2}{{){{{E_{b}(x_{m},t}}_{3}{)}%
E_{b}(x_{m},t}}}_{4}{{)}}\\
-%
{\displaystyle\sum\limits_{n}}
i\left(  \frac{2{d}_{0}^{2}}{c^{2}}\right)  ^{2}\int dt_{1}dt_{2}dt_{3}%
dt_{4}\cdot{G}_{e}\left(  t_{1}-t_{2}\right)  {G}_{g}\left(  t_{2}%
-t_{3}\right)  {G}_{e}\left(  t_{3}-t_{4}\right) \\
\cdot{{{{E_{a}(x_{n},t}}_{1}{)}E_{a}(x_{n},t}}_{2}{{){{E_{b}(x_{n},t}}%
_{3}{{){{E_{b}(x_{n},t}}_{4}{)}}}}}%
\end{array}
\right)  \label{eq_integ}%
\end{equation}
If we introduce the atom number density as
\[
\rho\left(  x\right)  =\sum\limits_{n}\delta(x-x_{n})
\]
The discrete description of atomic ensemble can be replaced by continuous one
and equation \ref{eq_integ} transforms into%
\begin{equation}
\prod\limits_{n}Z_{n}=\mathcal{\breve{N}}\left(
\begin{array}
[c]{c}%
1-i\frac{2{d}_{0}^{2}}{c^{2}}\int dt_{1}dt_{2}d^{3}x\cdot\rho\left(  x\right)
{{{{E_{a}(x,t}}_{1}{)}E_{a}(x,t}}_{2}{)\cdot G}_{e}\left(  t_{1}-t_{2}\right)
\\
+\frac{1}{2}\left(  -i\frac{2{d}_{0}^{2}}{c^{2}}\right)  ^{2}\int dt_{1}%
dt_{2}dt_{3}dt_{4}{{\cdot d^{3}x{d^{3}y}\cdot}}\left[  {{{{\rho\left(
x\right)  }\rho\left(  y\right)  -{\rho\left(  x\right)  \delta}}}}^{3}\left(
x-y\right)  \right] \\
{\cdot G}_{e}\left(  t_{1}-t_{2}\right)  {G}_{e}\left(  t_{3}-t_{4}\right)
{{{{E_{a}(x,t}}_{1}{)}E_{a}(x,t}}_{2}{{){{{E_{b}(y,t}}_{3}{)}E_{b}(y,t}}}%
_{4}{{)}}\\
-i\left(  \frac{2{d}_{0}^{2}}{c^{2}}\right)  ^{2}\int dt_{1}dt_{2}dt_{3}%
dt_{4}d^{3}x\cdot\rho\left(  x\right)  \cdot{G}_{e}\left(  t_{1}-t_{2}\right)
{G}_{g}\left(  t_{2}-t_{3}\right)  {G}_{e}\left(  t_{3}-t_{4}\right) \\
\cdot{{{{E_{a}(x,t}}_{1}{)}E_{a}(x,t}}_{2}{{){{E_{b}(x,t}}_{3}{{){{E_{b}(x,t}%
}_{4}{)}}}}}%
\end{array}
\right)
\end{equation}

The dipole density $\rho(x)$ is stochastic value and depends on realization of
disorder in the investigated sample. Unless the problem under consideration is
light propagation in a certain sample, the answer has to averaged over density
realization. Denote by $\mathbf{P}\left[  \rho(x)\right]  $ the probability
density that the atom number density is equal to $\rho(x)$. The rigorous
definition of this density reads%
\[
\mathbf{P}\left[  \rho(x)\right]  =C\exp\left[  L\left[  \rho(x)\right]
\right]
\]
where the normalization constant is defined as%

\[
C^{-1}=\int D\rho(x)\exp\left[  L\left[  \rho(x)\right]  \right]
\]

Generally, realization of disorder may vary with time (it means, in the course
of evolution the corresponding cumulants change their values), but we restrict
our investigation by the frozen disorder, where the first and second cumulant
are not equal to zero. This approximation holds true for low density, when a
typical contributions of the higher cumulants drops remarkably with increasing
its order\cite{Ziman79Disoder}, so $L\left[  \rho(x)\right]  $ can expressed as%

\[
L\left[  \rho(x)\right]  =-\frac{1}{2}\int dx_{1}dx_{2}\frac{\left(
\rho(x_{1})-\rho_{0}(x_{1})\right)  \left(  \rho(x_{2})-\rho_{0}%
(x_{2})\right)  }{\Gamma^{(2)}(x_{1},x_{2})}%
\]
Moreover, the atomic gas is assumed to be translation invariant and the
density-density correlation can be expressed as
\begin{align*}
\Gamma^{(2)}(x,y)  &  =\left\langle {\rho(x)\rho(y)}\right\rangle -\rho
_{0}^{2}=\rho_{0}^{2}C(x-y)+\rho_{0}\delta^{3}(x-y)\\
\left\langle \rho(x)\right\rangle  &  =\rho_{0}%
\end{align*}
where $C(x-y)$ is dimensionless correlation function for density,(
density-density correlation is supposed to be independent on the atoms
positions and interatomic distance affects on the the correlation function ).
So averaging over position we get eventually
\begin{align}
\prod\limits_{n}Z_{n}  &  =\mathcal{\breve{N}K}\left(  t,E,C,\rho_{0}\right)
\\
&  =\mathcal{\breve{N}}\left(
\begin{array}
[c]{c}%
1-i\frac{2{d}_{0}^{2}\rho_{0}}{c^{2}}\int dt_{1}dt_{2}d^{3}x\cdot
{{{{E_{a}(x,t}}_{1}{)}E_{a}(x,t}}_{2}{)\cdot G}_{e}\left(  t_{1}-t_{2}\right)
\\
+\frac{1}{2}\left(  -i\frac{2{d}_{0}^{2}\rho_{0}}{c^{2}}\right)  ^{2}\int
dt_{1}dt_{2}dt_{3}dt_{4}{{\cdot d^{3}x{d^{3}y}}}\\
{\cdot G}_{e}\left(  t_{1}-t_{2}\right)  {G}_{e}\left(  t_{3}-t_{4}\right)
{{{{E_{a}(x,t}}_{1}{)}E_{a}(x,t}}_{2}{{){{{E_{b}(y,t}}_{3}{)}E_{b}(y,t}}}%
_{4}{{)}}\\
+\frac{1}{2}\left(  -i\frac{2{d}_{0}^{2}\rho_{0}}{c^{2}}\right)  ^{2}\int
dt_{1}dt_{2}dt_{3}dt_{4}{{\cdot d^{3}x{d^{3}y}\cdot}}C(x-y)\\
{\cdot G}_{e}\left(  t_{1}-t_{2}\right)  {G}_{e}\left(  t_{3}-t_{4}\right)
{{{{E_{a}(x,t}}_{1}{)}E_{a}(x,t}}_{2}{{){{{E_{b}(y,t}}_{3}{)}E_{b}(y,t}}}%
_{4}{{)}}\\
-i\left(  \frac{2{d}_{0}^{2}}{c^{2}}\right)  ^{2}\int dt_{1}dt_{2}dt_{3}%
dt_{4}d^{3}x\cdot\rho_{0}\cdot{G}_{e}\left(  t_{1}-t_{2}\right)  {G}%
_{g}\left(  t_{2}-t_{3}\right)  {G}_{e}\left(  t_{3}-t_{4}\right) \\
\cdot{{{{E_{a}(x,t}}_{1}{)}E_{a}(x,t}}_{2}{{){{E_{b}(x,t}}_{3}{{){{E_{b}(x,t}%
}_{4}{)}}}}}%
\end{array}
\right)
\end{align}
and the partition function is given by%

\[
Z_{1}=\mathcal{N}\int\mathcal{D}\left\{  {A^{0}A^{1}A^{2}A^{3}}\right\}
\exp[iS_{field}^{1}+\log\mathcal{K}\left(  t,E,C,\rho_{0}\right)  ]
\]

\section{Propagator for the electric field}

In the previous section we have deduced the partition function for the atomic
ensemble. On the basis of it we can establish a character of evolution for
4-th potential $A_{i}\left(  xt\right)  $, characterized the electromagnetic
field evolution. In itself the 4-th potential is not physical observable
value, so we focus our attention on electric field and calculate the
propagators for the electric field:
\[
D_{ab}({R}_{1}{t_{1,}{{R}_{2}t_{2})}}=-i\left\langle {{E_{a}({R}_{1}}%
t_{1}{)E_{b}({R}_{2}t_{2})}}\right\rangle
\]
The propagators of Greens function carry out complete information about
spectrum or dynamics of electric fields. It describes the propagation of waves
form an external $\delta$-delta function source. For further derivation it is
convenient to express the action in terms of the electric field. Namely, the
4-vector $A_{\alpha}$ can be expressed in terms of 4-vector included
components of electric field strength $E_{\alpha}(x\tau)$ supplemented with
zero component $E_{0}(x\tau)=-\partial_{0}A_{0}-\partial_{\alpha}A_{\alpha}%
$.Hence, the action of free electromagnetic field\ref{Action_field} in
impulse-frequency representation converts into%
\begin{align*}
S_{field}^{1}  &  =\hfill\frac{1}{{8\pi c}}\int\frac{d^{3}k}{\left(
2\pi\right)  ^{3}}\frac{d{{{{\Omega}}}}}{2\pi}{\left[  {\ -\frac{1}{\xi}%
E_{0}(k,{{{\Omega}}})E_{0}(-k,-{{{\Omega}}})+E_{a}(k{{{\Omega}}}%
)E_{b}(-k,-{{{\Omega}}})}\Pi_{ab}(k,{{{{\Omega}}}})\right]  }\\
\Pi_{ab}(k,{{{{\Omega}}}})  &  ={\frac{{k_{a}k_{b}}}{{{{{\Omega}}}^{2}}%
}+\delta_{ab}\frac{{{{{\Omega}}}^{2}}-k_{c}^{2}}{{{{{\Omega}}}^{2}}}}%
\end{align*}
The zero component of 4-vector $\left(  E_{0},E_{\alpha}\right)
,\alpha=x,y,z$ corresponds to the calibration and does not influence on any
physical observable values. This fact is proved by explicit representation of
$S_{field}^{1}$, and $\mathcal{K}\left(  t,E,C,\rho_{0}\right)  $ does not
contain $E_{0}$.

\begin{figure}[ptb]
\includegraphics[width=\textwidth]
{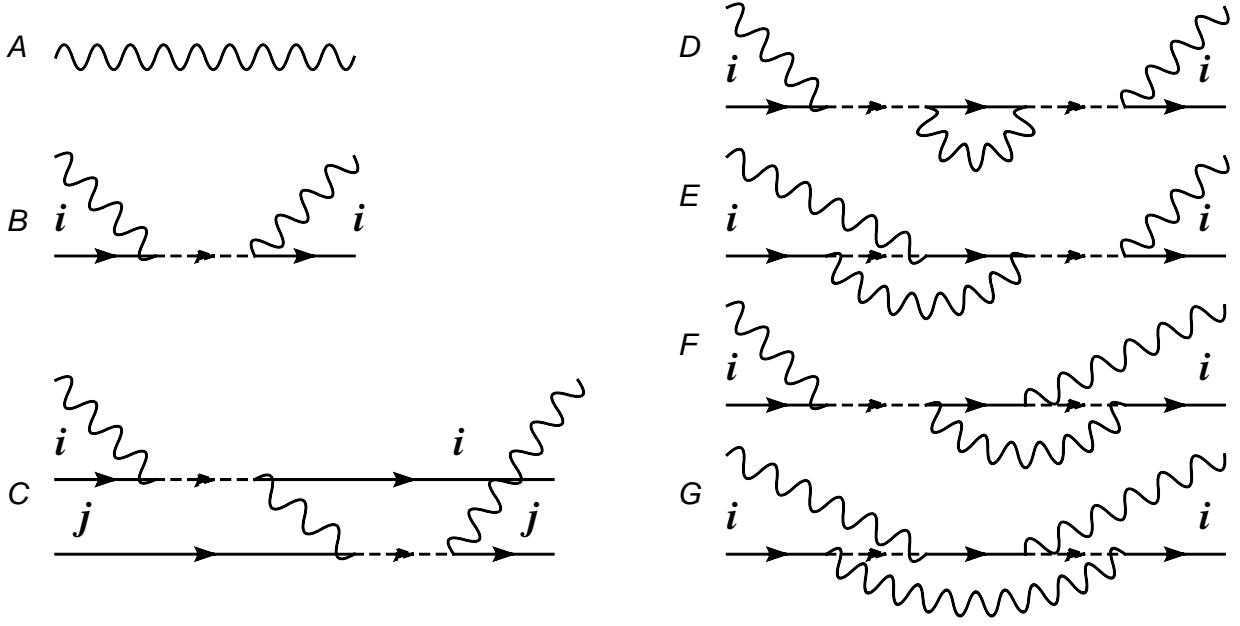}\newline \caption{Diagram representation for the photon Greens
function. The wavy, solid and dash lines correspond to free photon propagator,
the ground and excited atom Green functions respectively. }%
\label{fig1}%
\end{figure}

The Greens function of the photon state is then determined by
\begin{align*}
D_{ab}({R}_{1}{t_{1,}{{R}_{2}t_{2})}}  &  {=-i}{}\left\langle {E_{a}(R}%
_{1}{,T}_{1}{)E_{b}(R}_{2}T_{2}{)}\right\rangle \\
&  =-i\frac{\mathcal{\breve{N}}\int\prod\limits_{k=0,1,2,3}\mathcal{D}%
E_{k}\exp\left[  iS_{field}^{1}\left(  E_{k}\right)  \right]  \mathcal{K}%
\left(  E_{k},t\right)  \cdot{E_{a}(R}_{1}{,T}_{1}{)E_{b}(R}_{2}T_{2}{)}%
}{\mathcal{\breve{N}}\int\prod\limits_{k=0,1,2,3}\mathcal{D}E_{k}\exp\left[
iS_{field}^{1}\left(  E_{k}\right)  \right]  \mathcal{K}\left(  E_{k}%
,t\right)  }%
\end{align*}
this formula can be represented diagrammatically(see fig \ref{fig1}).

The right hand side features the contributions of free evolution (diagram A),
single scattering (diagram B), two scattering on two atoms (diagram A) and
double scattering on the same atom (diagrams D-G). The explicit expression of
the free photon propagator is%

\begin{align*}
D_{ab}^{0}({R}_{1}T{_{1,}{{R}_{2}T_{2})}} &  {=-i}{}\left\langle {E_{a}(R}%
_{1}{,T}_{1}{)E_{b}(R}_{2}T_{2}{)}\right\rangle \\
&  =\int\frac{{d}^{3}{kd\Omega}}{\left(  2\pi\right)  ^{4}}\exp\left[
{-i\Omega}\left(  {T}_{2}-T_{1}\right)  {+ik}_{c}\left(  {R}_{2c}%
-R_{1c}\right)  \right]  \frac{{4\pi c}}{{\Omega}^{2}-{k^{2}+i\varepsilon}%
}\cdot\left(  {\Omega{^{2}}}\delta_{ab}-{{{{k_{a}{k_{b}}}}}}\right)
\end{align*}
while the first order contribution is given by:%
\begin{align*}
D_{ab}^{1}({R}_{1}T{_{1,}{{R}_{2}T_{2})}}  & =\frac{2{d}_{0}^{2}\rho_{0}%
}{c^{2}}\int dt_{1}dt_{2}d^{3}x{\cdot}\left[  {G}_{e}\left(  t_{1}%
-t_{2}\right)  +{G}_{e}\left(  t_{2}-t_{1}\right)  \right]  \\
& \cdot D_{ap}^{0}({R}_{1}T{_{1,}x}t{_{1}{)}}D_{pb}^{0}({x}t{_{2,}{{R}%
_{2}T_{2})}}%
\end{align*}
The second order term contains two different contributions. The first one
appears as the second order Born term for the one-atom scattering and can be
presented as:%

\begin{align*}
D_{ab}^{2,1}({R}_{1}T{_{1,}{{R}_{2}T_{2})}} &  =i\left(  \frac{2{d}_{0}^{2}%
}{c^{2}}\right)  ^{2}\rho_{0}\int dt_{1}dt_{2}dt_{3}dt_{4}d^{3}x\\
&  \cdot D_{pa}^{0}\left(  R_{1}-x,T_{1}-t_{1}\right)  D_{bq}^{0}\left(
x-R_{2},t_{2}-T_{2}\right)  \cdot\frac{D_{bb}^{0}\left(  0,t_{4}-t_{3}\right)
}{3}\\
&  \left(
\begin{array}
[c]{c}%
+{G}_{e}\left(  t_{4}-t_{1}\right)  \left[  {G}_{g}\left(  t_{1}-t_{3}\right)
{G}_{e}\left(  t_{3}-t_{2}\right)  +{G}_{g}\left(  t_{1}-t_{2}\right)  {G}%
_{e}\left(  t_{2}-t_{3}\right)  \right]  \\
+{G}_{e}\left(  t_{1}-t_{4}\right)  \left[  {G}_{g}\left(  t_{4}-t_{3}\right)
{G}_{e}\left(  t_{3}-t_{2}\right)  +{G}_{g}\left(  t_{4}-t_{2}\right)  {G}%
_{e}\left(  t_{2}-t_{3}\right)  \right]  \\
+{G}_{e}\left(  t_{4}-t_{2}\right)  \left[  {G}_{g}\left(  t_{2}-t_{3}\right)
{G}_{e}\left(  t_{3}-t_{1}\right)  +{G}_{g}\left(  t_{2}-t_{1}\right)  {G}%
_{e}\left(  t_{1}-t_{3}\right)  \right]  \\
+{G}_{e}\left(  t_{2}-t_{4}\right)  \left[  {G}_{g}\left(  t_{4}-t_{3}\right)
{G}_{e}\left(  t_{3}-t_{1}\right)  +{G}_{g}\left(  t_{4}-t_{1}\right)  {G}%
_{e}\left(  t_{1}-t_{3}\right)  \right]
\end{array}
\right)
\end{align*}
The other term is in charge for the consequences photon scattering on two
different atoms.So, in contrast to the previous the second order term it is
proportional to the square of atom density and is equal to
\begin{align*}
D_{ab}^{2,2}({R}_{1}T{_{1,}{{R}_{2}T_{2})}} &  =\left(  \frac{2{d}_{0}^{2}%
\rho_{0}}{c^{2}}\right)  ^{2}\int dt_{1}dt_{2}dt_{3}dt_{4}{{\cdot d^{3}%
x{d^{3}y}\cdot}}\left[  1+\frac{C(x-y)+C(y-x)}{2}\right]  \\
&  {\cdot}\left(  {G}_{e}\left(  t_{4}-t_{1}\right)  +{G}_{e}\left(
t_{1}-t_{4}\right)  \right)  \left(  {G}_{e}\left(  t_{2}-t_{3}\right)
+{G}_{e}\left(  t_{3}-t_{2}\right)  \right)  \\
&  \cdot D_{pq}^{0}\left(  x-y,t_{4}-t_{3}\right)  \cdot D_{ap}^{0}\left(
R_{1}-x,T_{1}-t_{1}\right)  D_{bq}^{0}\left(  y-R_{2},t_{2}-T_{2}\right)
\end{align*}
The straightforward result can be obtained in the momentum-frequency
representation, where the free propagator is given by%

\[
D_{ab}^{0}\left(  k,\omega\right)  =\frac{{4\pi c}}{\omega{{^{2}}}%
-{{{{k^{2}+i\varepsilon}}}}}\cdot\left(  \omega{{^{2}}}\delta_{ab}%
-{{{{k_{a}{k_{b}}}}}}\right)
\]
The terms $D_{ab}^{2,2}$ and $D_{ab}^{1}$ can be easily calculated and are
equal to%
\begin{align*}
D_{ab}^{1}\left(  k,\Omega\right)   &  ={D}_{ap}^{0}\left(  k,\Omega\right)
\cdot\chi\left(  \Omega\right)  \cdot{D}_{pb}^{0}\left(  k,\Omega\right) \\
D_{ab}^{2,2}\left(  k,\Omega\right)   &  ={D}_{ap}^{0}\left(  k,\Omega\right)
\cdot\chi\left(  \Omega\right)  \left[  D_{pq}^{0}\left(  k,\Omega\right)
+H_{pq}\left(  k,\Omega\right)  \right]  \chi\left(  \Omega\right)  \cdot
{D}_{qb}^{0}\left(  k,\Omega\right) \\
\chi\left(  \Omega\right)   &  =\frac{4\Delta{d}_{0}^{2}\rho_{0}}{c^{2}}%
\frac{1}{\Omega^{2}-\Delta^{2}+i\delta}\\
H_{ab}\left(  k,\Omega\right)   &  =%
{\displaystyle\int}
\frac{d^{3}k^{\prime}}{\left(  2\pi\right)  ^{3}}{D}_{ab}\left(  k^{\prime
},\Omega\right)  \cdot\frac{C\left(  k^{\prime}-k\right)  +C\left(  k^{\prime
}+k\right)  }{2}%
\end{align*}
The expression for $D_{ab}^{2,1}$ after the Fourier transformation takes the form%

\begin{align*}
D_{ab}^{2,1}\left(  k,\Omega\right)   &  =i\left(  \frac{2{d}_{0}^{2}}{c^{2}%
}\right)  ^{2}\rho_{0}B\left(  \Omega\right)  {D}_{ap}^{0}\left(
k,\Omega\right)  {D}_{pb}^{0}\left(  k,\Omega\right) \\
B\left(  \Omega_{1}\right)   &  =%
{\displaystyle\int}
\frac{d^{3}k}{\left(  2\pi\right)  ^{3}}\frac{d\Omega}{2\pi}\frac{1}{3}%
D_{bb}^{0}\left(  k,\Omega\right)  \left[  V\left(  \Omega_{1},\Omega\right)
+V\left(  -\Omega_{1},\Omega\right)  \right] \\
V\left(  \Omega_{1},\Omega\right)   &  =G_{g}\left(  \Omega_{1}-\Omega\right)
\left(  G_{e}\left(  -\Omega\right)  +G_{e}\left(  \Omega_{1}\right)  \right)
^{2}%
\end{align*}
The integral for $B\left(  \Omega_{1}\right)  $ is divergent and the
singularities are connected to the large-k behavior of the Green's function in
Fourier space. So, in order to remove the singularities, one can modify this
behavior. The standard method, often used in quantum field theory, is to
introduce a large-momentum cutoff. Because of the vector character of the
photon propagator, we allow for two regularization procedures with different
parameters, $\Lambda_{L}$ and $\Lambda_{T}$%

\begin{align*}
D_{aa}^{0}(0,{{{{{\Omega}}}}})  &  =%
{\displaystyle\int}
\frac{d^{3}k}{\left(  2\pi\right)  ^{3}}D_{bb}^{0}\left(  k,{{{{{\Omega}}}}%
}\right) \\
&  =%
{\displaystyle\int}
\frac{d^{3}k}{\left(  2\pi\right)  ^{3}}\left[  D_{bb}^{0}\left(
k,{{{{{\Omega}}}}}\right)  -{4\pi c}\left(  1-\frac{2{{{{{\Omega}}}^{2}}}%
}{k^{2}}\right)  \right]  +%
{\displaystyle\int}
\frac{d^{3}k}{\left(  2\pi\right)  ^{3}}{4\pi c}\left(  1-\frac{2{{{{{\Omega}%
}}^{2}}}}{k^{2}}\right) \\
&  =c\Lambda_{L}^{3}-c{{{{{\Omega}}}}^{2}}\Lambda_{T}-i{2c}\left\vert
{{{{{\Omega}}}}}^{3}\right\vert
\end{align*}
The employed regularization procedure is of course not unique, but it suffices
for our purposes.The physical result is largely insensitive to such details.
The final expression for $D_{ab}^{2,1}\left(  k,\Omega\right)  $%

\begin{align*}
D_{ab}^{2,1}\left(  k,\Omega\right)   &  =i\left(  \frac{2{d}_{0}^{2}}{c^{2}%
}\right)  ^{2}\rho_{0}B\left(  \Omega\right)  {D}_{ap}^{0}\left(
k,\Omega\right)  {D}_{pb}^{0}\left(  k,\Omega\right) \\
B\left(  \Omega\right)   &  =-iC_{1}\left[  \frac{2\Delta}{\Omega^{2}%
-\Delta^{2}+i\delta}\right]  -iC_{2}\left[  \frac{2\Delta}{\Omega^{2}%
-\Delta^{2}+i\delta}\right]  ^{2}-{c}\frac{2\left\vert \Omega^{3}\right\vert
}{3}\left[  \frac{2\Delta}{\Omega^{2}-\Delta^{2}+i\delta}\right]  ^{2}%
\end{align*}
where the constant $C_{1}$ and $C_{2}$ are related with the cutoff parameter.
The total propagator is given by:%
\begin{align}
{D}_{ab}\left(  k,\Omega\right)   &  ={D}_{ab}^{0}\left(  k,\Omega\right)
+{D}_{ap}^{0}\left(  k,\Omega\right)  \cdot\chi\left(  \Omega\right)  \cdot
{D}_{pb}^{0}\left(  k,\Omega\right) \label{GenAnswer}\\
&  +{D}_{ap}^{0}\left(  k,\Omega\right)  \cdot\chi\left(  \Omega\right)
\left[  D_{pq}^{0}\left(  k,\Omega\right)  +H_{pq}\left(  k,\Omega\right)
\right]  \chi\left(  \Omega\right)  \cdot{D}_{qb}^{0}\left(  k,\Omega\right)
\nonumber\\
&  +{D}_{ap}^{0}\left(  k,\Omega\right)  \chi\left(  \Omega\right)
\cdot\left(  -{ci}\frac{2\left\vert \Omega^{3}\right\vert }{3\rho_{0}}%
\chi\left(  \Omega\right)  \delta_{pq}\right)  \cdot{D}_{qb}^{0}\left(
k,\Omega\right) \nonumber\\
\chi\left(  \Omega\right)   &  =\frac{4\Delta{d}_{0}^{2}\rho_{0}}{c^{2}}%
\frac{1}{\Omega^{2}-\Delta^{2}+i\delta}\nonumber\\
H_{ab}\left(  k,\Omega\right)   &  =%
{\displaystyle\int}
\frac{d^{3}k^{\prime}}{\left(  2\pi\right)  ^{3}}{D}_{ab}\left(  k^{\prime
},\Omega\right)  \cdot\frac{C\left(  k^{\prime}-k\right)  +C\left(  k^{\prime
}+k\right)  }{2}%
\end{align}
We have omitted the terms included renormalization constants $C_{1},C_{2}$,
because they result in shifting of the values of dipole momentum $d_{0}$ and
energy $\Delta$.

To specify the problem completely one needs to give the value for the
density-density correlation $C\left(  r\right)  $. We start consideration from
white-noise correlation, where $C\left(  r\right)  =K\delta^{3}\left(
r\right)  $. Such a function describes short-range correlation appears between
the neighboring atoms. With accuracy of renormalization constant
\[
H_{ab}\left(  k,\Omega\right)  =-{2}i{c}\left\vert {{{{{\Omega}}}}}%
^{3}\right\vert K
\]
and%
\begin{align*}
{D}_{ab}\left(  k,\Omega\right)   &  ={D}_{ab}^{0}\left(  k,\Omega\right)
+{D}_{ap}^{0}\left(  k,\Omega\right)  \cdot\chi\left(  \Omega\right)  \cdot
{D}_{pb}^{0}\left(  k,\Omega\right) \\
&  +{D}_{ap}^{0}\left(  k,\Omega\right)  \cdot\chi\left(  \Omega\right)  \cdot
D_{pq}^{0}\left(  k,\Omega\right)  \cdot\chi\left(  \Omega\right)  \cdot
{D}_{qb}^{0}\left(  k,\Omega\right) \\
&  +{D}_{ap}^{0}\left(  k,\Omega\right)  \chi\left(  \Omega\right)  \cdot
\frac{-2{ci}\left\vert \Omega^{3}\right\vert }{3\rho_{0}}\left(  1+K\rho
_{0}\right)  \chi\left(  \Omega\right)  \cdot{D}_{pb}^{0}\left(
k,\Omega\right) \\
\chi\left(  \Omega\right)   &  =\frac{4\Delta{d}_{0}^{2}\rho_{0}}{c^{2}}%
\frac{1}{\Omega^{2}-\Delta^{2}+i\delta}%
\end{align*}
For the low density of scatters with the same accuracy the answer can be
present in a compact form%

\begin{align*}
{D}_{ab}\left(  k,\Omega\right)   &  ={D}_{ab}^{0}\left(  k,\Omega\right)
+{D}_{ap}^{0}\left(  k,\Omega\right)  \cdot\chi\left(  \Omega\right)  \cdot
D_{pq}^{0}\left(  k,\Omega\right)  \cdot\chi\left(  \Omega\right)  \cdot
{D}_{qb}^{0}\left(  k,\Omega\right)  \\
&  +{D}_{ap}^{0}\left(  k,\Omega\right)  \cdot\frac{1}{\left[  \chi\left(
\Omega\right)  \right]  ^{-1}}\left[  1-\frac{2{ci}\left\vert \Omega
^{3}\right\vert \left(  1+K\rho_{0}\right)  }{3\rho_{0}\left[  \chi\left(
\Omega\right)  \right]  ^{-1}}\right]  \cdot{D}_{pb}^{0}\left(  k,\Omega
\right)  \\
&  \simeq{D}_{ab}^{0}\left(  k,\Omega\right)  +{D}_{ap}^{0}\left(
k,\Omega\right)  \cdot\tilde{\chi}\left(  \Omega\right)  \cdot{D}_{pb}%
^{0}\left(  k,\Omega\right)  \\
&  +{D}_{ap}^{0}\left(  k,\Omega\right)  \cdot\tilde{\chi}\left(
\Omega\right)  \cdot{D}_{pq}^{0}\left(  k,\Omega\right)  \cdot\tilde{\chi
}\left(  \Omega\right)  \cdot{D}_{qb}^{0}\left(  k,\Omega\right)  \\
\tilde{\chi}\left(  \Omega\right)   &  =\frac{4\Delta{d}_{0}^{2}\rho_{0}%
}{c^{2}}\frac{1}{\Omega^{2}-\Delta^{2}+\frac{8i}{3c}\Delta{d}_{0}%
^{2}\left\vert \Omega^{3}\right\vert \left(  1+K\rho_{0}\right)  }%
\end{align*}
or%
\begin{align*}
{D}_{ab}\left(  \Omega,k\right)   &  ={4\pi c}\frac{\delta_{ab}\Omega
^{2}\left(  1{-}\alpha\left(  \Omega\right)  \right)  -{{{k}}}_{a}k_{b}%
}{\Omega^{2}\left(  1{-}\alpha\left(  \Omega\right)  \right)  -{k^{2}%
+i\varepsilon}}\frac{1}{\left(  1{-}\alpha\left(  \Omega\right)  \right)  }\\
\alpha\left(  \Omega\right)   &  =\frac{16\pi\Delta{d}_{0}^{2}\rho_{0}}%
{c}\frac{1}{\Omega^{2}-\Delta^{2}+\frac{8i{d}_{0}^{2}\Delta\left\vert
\Omega^{3}\right\vert }{3c}\left(  1+K\rho_{0}\right)  }%
\end{align*}
Here $\alpha\left(  \Omega\right)  $ coincides with permittivity for the
dipole gas. Because of the correlation the imaginary part of permittivity has
increased in contrast to the common dipolar gas. Such a behavior leads to
increasing the scattering mean free path $l_{R}$. It is determined by the
imaginary part of the pole%

\[
\frac{1}{2l_{R}}=\operatorname{Im}\sqrt{\Omega^{2}\left(  1{-}\alpha\left(
\Omega\right)  \right)  }%
\]
The result reads%
\begin{align*}
l_{R}  &  \simeq\frac{2}{3\pi}\frac{\left\vert \Omega^{2}\right\vert \left(
1+K\rho_{0}\right)  }{\rho_{0}}\frac{\left(  \Omega^{2}-\Delta^{2}\right)
^{2}+\left(  \Gamma\left\vert \Omega^{3}\right\vert \right)  ^{2}}{\left(
\Gamma\left\vert \Omega^{3}\right\vert \right)  ^{2}}\\
\Gamma &  =\frac{8{d}_{0}^{2}\Delta}{3c}\left(  1+K\rho_{0}\right)
\end{align*}
As expected, this result indicates that with increasing correlation scattering
is determined by multiply scattering, while for the low values $K$ the
imaginary part is determined by the evolution from the internal degrees.

Above we have analyzed the short-correlation case, but the situation becomes
drastically different when long range correlation sets in due to induced
particle aggregation, as for instance in gelly or aerogels. If it is the case
the series\ref{GenAnswer} can collected as follows:%

\begin{align}
{D}_{ab}\left(  k,\Omega\right)   &  \simeq{D}_{ap}^{0}\left(  k,\Omega
\right)  +{D}_{ap}^{0}\left(  k,\Omega\right)  \tilde{\chi}\left(
\Omega\right)  {D}_{pb}^{0}\left(  k,\Omega\right) \nonumber\\
&  +{D}_{ap}^{0}\left(  k,\Omega\right)  \cdot\tilde{\chi}\left(
\Omega\right)  \left[  D_{pq}^{0}\left(  k,\Omega\right)  +H_{pq}\left(
k,\Omega\right)  \right]  \tilde{\chi}\left(  \Omega\right)  \cdot{D}_{qb}%
^{0}\left(  k,\Omega\right) \nonumber\\
\tilde{\chi}\left(  \Omega\right)   &  =\frac{4\Delta{d}_{0}^{2}\rho_{0}%
}{c^{2}}\frac{1}{\Omega^{2}-\Delta^{2}+\frac{8i}{3c}\Delta{d}_{0}%
^{2}\left\vert \Omega^{3}\right\vert }%
\end{align}
Thus, with the accuracy upto the second order the electric field propagator
can be rewritten as%

\begin{align}
{D}_{ab}\left(  k,\Omega\right)   &  \simeq{D}_{ab}^{eff}\left(
k,\Omega\right)  +{D}_{ap}^{eff}\left(  k,\Omega\right)  \cdot\tilde{\chi
}\left(  \Omega\right)  H_{pq}\left(  k,\Omega\right)  \tilde{\chi}\left(
\Omega\right)  \cdot{D}_{pb}^{eff}\left(  k,\Omega\right) \nonumber\\
{D}_{ab}^{eff}\left(  k,\Omega\right)   &  ={4\pi c}\frac{\delta_{ab}%
\Omega^{2}\left(  1{-4\pi c}\tilde{\chi}\left(  \Omega\right)  \right)
-{{{k}}}_{a}k_{b}}{\Omega^{2}\left(  1{-4\pi c}\tilde{\chi}\left(
\Omega\right)  \right)  -{k^{2}+i\varepsilon}}\frac{1}{\left(  1{-4\pi
c}\tilde{\chi}\left(  \Omega\right)  \right)  }%
\end{align}
The first term of the Green function can be attribute to a photon free
propagation in the atomic resonance gas without correlation, while the second
term is related with "scattering on density correlation" in such a medium.
Taking into account, that the medium is translation invariant, one can
decompose $H_{ab}\left(  k,\Omega\right)  $ as follows%

\begin{equation}
H_{ab}\left(  k,\Omega\right)  =%
{\displaystyle\int}
\frac{d^{3}k^{\prime}}{\left(  2\pi\right)  ^{3}}{D}_{ab}\left(  k^{\prime
},\Omega\right)  \cdot\frac{C\left(  k^{\prime}-k\right)  +C\left(  k^{\prime
}+k\right)  }{2}=\tilde{A}\left(  k,\Omega\right)  \delta_{ab}+\tilde
{B}\left(  k,\Omega\right)  \frac{k_{a}k_{b}}{k^{2}} \label{Hab}%
\end{equation}
Thus, in contrast to the short-range correlation, the long-range density
correlations result in space anisotropy for the light propagation. The
character of this anisotropy is determined by the coefficients $\tilde
{A}\left(  k,\Omega\right)  ,\tilde{B}\left(  k,\Omega\right)  $. The function
$H_{ab}\left(  k,\Omega\right)  $ can be rewritten as
\begin{align*}
H_{ab}\left(  k,\Omega\right)   &  =-%
{\displaystyle\int}
{c}C\left(  R\right)  \cos\left[  \mathbf{kR}\right]  \cdot\left(  \delta
_{ab}\Omega^{2}+{{{\partial}}}_{a}\partial_{b}\right)  \frac{\exp\left[
i\left\vert \Omega\right\vert R\right]  }{R}\cdot d^{3}R\\
&  =-%
{\displaystyle\int}
{c}C\left(  R\right)  \cos\left[  \mathbf{kR}\right]  \cdot\left(  \delta
_{ab}\Omega^{2}+\delta_{ab}\frac{\Delta}{3}+{{{\partial}}}_{a}\partial
_{b}-\delta_{ab}\frac{\Delta}{3}\right)  \frac{\exp\left[  i\left\vert
\Omega\right\vert R\right]  }{R}\cdot d^{3}R
\end{align*}
The tensor $H_{ab}\left(  k,\Omega\right)  $ can be decomposed into terms:%
\begin{align*}
H_{ab}\left(  k,\Omega\right)   &  =H_{ab}^{1}\left(  k,\Omega\right)
+H_{ab}^{2}\left(  k,\Omega\right) \\
H_{ab}^{1}\left(  k,\Omega\right)   &  =-\int{c}C\left(  R\right)  \cos\left[
\mathbf{kR}\right]  \cdot\left(  {{{\partial}}}_{a}\partial_{b}-\delta
_{ab}\frac{\Delta}{3}\right)  \frac{\exp\left[  i\left\vert \Omega\right\vert
R\right]  }{R}\cdot d^{3}R\\
H_{ab}^{2}\left(  k,\Omega\right)   &  =-\int{c}C\left(  R\right)  \cos\left[
\mathbf{kR}\right]  \cdot\delta_{ab}\left(  \Omega^{2}+\frac{\Delta}%
{3}\right)  \frac{\exp\left[  i\left\vert \Omega\right\vert R\right]  }%
{R}\cdot d^{3}R
\end{align*}
The trace for the first term is equal to 0, so a general expression for
$H_{ab}^{1}\left(  k,\Omega\right)  $%
\[
H_{ab}^{1}\left(  k,\Omega\right)  =A\left(  k,\Omega\right)  \left(
\delta_{ab}-3\frac{k_{a}k_{b}}{k^{2}}\right)
\]
and the value for $A\left(  k,\Omega\right)  $ can be found if we multiply the
tensor on ~$\frac{k_{a}k_{b}}{k^{2}}$, so%

\begin{align*}
\frac{k_{a}k_{b}}{k^{2}}H_{ab}^{1}\left(  k,\Omega\right)   &  =-2A\left(
k,\Omega\right)  \\
&  =-\frac{k_{a}k_{b}}{k^{2}}\int{c}C\left(  R\right)  \cos\left[
\mathbf{kR}\right]  \cdot\left(  \frac{R_{a}R_{b}}{R^{2}}-\frac{\delta_{ab}%
}{3}\right)  R\\
&  \cdot\frac{d}{dR}\left(  \frac{1}{R}\frac{d}{dR}\left(  \frac{\exp\left[
i\left\vert \Omega\right\vert R\right]  }{R}\right)  \right)  \cdot d^{3}R\\
&  =-\int{c}C\left(  R\right)  \cos\left[  kR\cos\theta\right]  \cdot\left(
\cos^{2}\theta-\frac{1}{3}\right)  R\\
&  \cdot\frac{d}{dR}\left(  \frac{1}{R}\frac{d}{dR}\left(  \frac{\exp\left[
i\left\vert \Omega\right\vert R\right]  }{R}\right)  \right)  \cdot2\pi
R^{2}dRd\cos\theta
\end{align*}
So after the simplification we get relationship for $A\left(  k,\Omega\right)
$%
\begin{align*}
P\left(  a\right)   &  =\frac{2\sin a}{a}\\
A\left(  k,\Omega\right)   &  =-{c}\pi\int C\left(  R\right)  \left(
P^{\prime\prime}\left(  kR\right)  +\frac{1}{3}P\left(  kR\right)  \right)
\cdot\frac{d}{dR}\left(  \frac{1}{R}\frac{d}{dR}\left(  \frac{\exp\left[
i\left\vert \Omega\right\vert R\right]  }{R}\right)  \right)  \cdot R^{3}dR
\end{align*}
The evolution the second term is simpler, because it is proportional to
$\delta_{ab}$. After the replacement Laplacian we get%

\begin{align*}
H_{ab}^{2}\left(  k,\Omega\right)   &  =-\delta_{ab}\int{c}C\left(  R\right)
\cos\left[  \mathbf{kR}\right]  \cdot\left(  \frac{2\Omega^{2}}{3}\frac
{\exp\left[  i\left\vert \Omega\right\vert R\right]  }{R}-\frac{4\pi}{3}%
\delta^{3}(R)\right)  \cdot d^{3}R\\
&  =\delta_{ab}\frac{4\pi{c}}{3}C\left(  0\right)  -\delta_{ab}\frac{8\pi
{c}\Omega^{2}}{3k}\int C\left(  R\right)  \cdot\exp\left[  i\left\vert
\Omega\right\vert R\right]  \sin kR\cdot dR
\end{align*}
The final expression for $H_{ab}\left(  k,\Omega\right)  $ is
\begin{align}
H_{ab}\left(  k,\Omega\right)   &  =\left[  A\left(  k,\Omega\right)
-B\left(  k,\Omega\right)  +\frac{4\pi{c}}{3}C\left(  0\right)  \right]
\delta_{ab}-3A\left(  k,\Omega\right)  \frac{k_{a}k_{b}}{k^{2}} \label{HabFin}%
\\
A\left(  k,\Omega\right)   &  =-{c}\pi\int C\left(  R\right)  \left(
P^{\prime\prime}\left(  kR\right)  +\frac{1}{3}P\left(  kR\right)  \right)
\cdot\frac{d}{dR}\left(  \frac{1}{R}\frac{d}{dR}\left(  \frac{\exp\left[
i\left\vert \Omega\right\vert R\right]  }{R}\right)  \right)  \cdot
R^{3}dR\nonumber\\
B\left(  k,\Omega\right)   &  =\frac{8\pi{c}\Omega^{2}}{3k}\int C\left(
R\right)  \cdot\exp\left[  i\left\vert \Omega\right\vert R\right]  \sin
kR\cdot dR\nonumber
\end{align}
From the structure of the expression(\ref{HabFin}) one can conclude, the only
long-range correlations cause space anisotropy for photon propagation.

\section{Conclusions}

We have studied the photon propagation through randomly
distributed resonant atoms. Starting from the exact action
included interatomic dynamics, atom-atom interaction(in
electric-dipole approximation) and electric field atom
interaction, we have derived the Green function for the dressed
photon that describes both the inelastic processes like a
spontaneous emission of the atoms and elastic scattering photon in
gas with density-density correlations via resonant atom-field
interaction. The proposed model of atoms allowed us to describe
consequence of transitions in real atoms and abandoned two-level
models \cite{Dikke} mostly used. The  applied approach are not
restricted to the rotating-wave approximation. We have shown that
the energy associated with transition between resonance states
does not change for low atomic density where the coupling
parameter is $\frac {16\pi{d}_{0}^{2}\rho_{0}}{\Delta}$ is small.
There are two energy-transfer channels in general: one channel
through resonant dipole-dipole interaction mediated by
virtual-photon creation and destruction and the other one through
emission and absorption of real photons. In particular for strong
atom-field interaction, the (over a period averaged) energy
transfer can be inhibited due to destructive interference of the
two available transfer channels. The developed approach allows to
consider both these channels simultaneously without separation.
Short-range and long-range correlation for the atom densities have
to be studied. From the general consideration, it was found that
even for translation invariant medium, long-range correlation
results to space anisotropy for the photon propagation. Clearly,
the present analysis has left a number of open questions, on which
future work will concentrate. In particular, the problem of energy
transfer in ensemble of the moving atoms or influence of
saturation effects on the character light propagation.

\begin{acknowledgments}
We wish to acknowledge J.T.M. Walraven for helpful discussions.
This work was supported by the Russian Foundation for Fundamental
Research(RFBR 05-02-17488-a), the Nederlandse Stichting voor
Fundamenteel Onderzoek der Materie (FOM). M.B.S. would also like
to thank Netherlands Organization for Scientific Research (NWO
047.019.005)
\end{acknowledgments}

\end{document}